\documentclass[sigconf]{acmart}

\AtBeginDocument{%
  }

\usepackage{listings}
\usepackage{svg}
\usepackage{algorithm, algpseudocode}
\usepackage{algorithmicx}
\usepackage{adjustbox}
\usepackage{multirow}
\usepackage{xspace}
\usepackage{xcolor}
\usepackage{colortbl}
\usepackage{soul}
\usepackage{amsthm}
\usepackage{lipsum}
\usepackage{pifont}
\usepackage{subcaption}
\usepackage{array}

\theoremstyle{definition}

\definecolor{verylightgray}{rgb}{.97,.97,.97}

\lstdefinelanguage{Solidity}{
	keywords=[1]{anonymous, assembly, assert, balance, break, call, callcode, case, catch, class, constant, continue, constructor, contract, debugger, default, delegatecall, delete, do, else, emit, event, experimental, export, external, false, finally, for, function, if, implements, import, in, indexed, instanceof, interface, internal, is, length, library, log0, log1, log2, log3, log4, memory, modifier, new, payable, pragma, private, protected, public, pure, push, require, return, returns, selfdestruct, send, solidity, storage, struct, suicide, super, switch, then, this, throw, transfer, true, try, typeof, using, value, view, while, with, ecrecover, keccak256, mulmod, ripemd160, sha256, sha3}, 
	keywordstyle=[1]\color{blue}\bfseries,
	keywords=[2]{address, bool, byte, bytes, bytes1, bytes2, bytes3, bytes4, bytes5, bytes6, bytes7, bytes8, bytes9, bytes10, bytes11, bytes12, bytes13, bytes14, bytes15, bytes16, bytes17, bytes18, bytes19, bytes20, bytes21, bytes22, bytes23, bytes24, bytes25, bytes26, bytes27, bytes28, bytes29, bytes30, bytes31, bytes32, enum, int, int8, int16, int24, int32, int40, int48, int56, int64, int72, int80, int88, int96, int104, int112, int120, int128, int136, int144, int152, int160, int168, int176, int184, int192, int200, int208, int216, int224, int232, int240, int248, int256, mapping, string, uint, uint8, uint16, uint24, uint32, uint40, uint48, uint56, uint64, uint72, uint80, uint88, uint96, uint104, uint112, uint120, uint128, uint136, uint144, uint152, uint160, uint168, uint176, uint184, uint192, uint200, uint208, uint216, uint224, uint232, uint240, uint248, uint256, var, void, ether, finney, szabo, wei, days, hours, minutes, seconds, weeks, years},	
	keywordstyle=[2]\color{teal}\bfseries,
	keywords=[3]{block, blockhash, coinbase, difficulty, gaslimit, number, timestamp, msg, data, sender, sig, value, now, tx, gasprice, origin},	
	keywordstyle=[3]\color{violet}\bfseries,
	identifierstyle=\color{black},
	sensitive=false,
	comment=[l]{//},
	morecomment=[s]{/*}{*/},
	commentstyle=\color{gray}\ttfamily,
	stringstyle=\color{red}\ttfamily,
	morestring=[b]',
	morestring=[b]"
}

\newcommand{\name}{\textsc{Proxion}\xspace}
\newcommand{\codename}{\name}
\newcommand{\delegate}{\texttt{DELEGATECALL}\xspace}

\newcommand{\para}[1]{\textbf {#1}.}

\newcommand{\minbox}[2]{%
  \mathmakebox[\ifdim#1<\width\width\else#1\fi]{#2}}
\newcommand{\Let}[2]{\State $ #1 \gets #2 $}

\newcommand{\cmark}{\ding{51}}%
\newcolumntype{P}[1]{>{\centering\arraybackslash}p{#1}}

\begin{document} 


\title{\name: Uncovering Hidden Proxy Smart Contracts for~Finding~Collision Vulnerabilities in Ethereum}


\author{Cheng-Kang Chen}
\authornote{Both authors contributed equally to this research.}
\affiliation{%
  \institution{National Taiwan University}
  \country{} 
}

\author{Wen-Yi Chu}
\authornotemark[1]
\affiliation{%
  \institution{National Taiwan University}
  \country{} 
}

\author{Muoi Tran}
\affiliation{%
  \institution{ETH Zürich}
  \country{} 
}

\author{Laurent Vanbever}
\affiliation{
  \institution{ETH Zürich}
  \country{} 
}

\author{Hsu-Chun Hsiao}
\affiliation{%
  \institution{National Taiwan University}
  \country{} 
}

\begin{abstract}

The proxy design pattern allows Ethereum smart contracts to be simultaneously immutable and upgradeable, in which an original contract is split into a proxy contract containing the data storage and a logic contract containing the implementation logic.
This architecture is known to have security issues, namely function collisions and storage collisions between the proxy and logic contracts, and has been exploited in real-world incidents to steal users' millions of dollars worth of digital assets.
In response to this concern, several previous works have sought to identify proxy contracts in Ethereum and detect their collisions. 
However, they all fell short due to their limited coverage, often restricting analysis to only contracts with available source code or past transactions.

To bridge this gap, we present \codename, an automated cross-contract analyzer that identifies all proxy smart contracts and their collisions in Ethereum. 
What sets \codename apart is its ability to analyze {\em hidden} smart contracts that lack both source code and past transactions. 
Equipped with various techniques to enhance efficiency and accuracy, \codename outperforms the state-of-the-art tools, notably identifying millions more proxy contracts and thousands of unreported collisions. 
We apply \codename to analyze over 36 million alive contracts from 2015 to 2023, revealing that 54.2\% of them are proxy contracts, and about 1.5 million contracts exhibit at least one collision issue.

\end{abstract}




\maketitle

\section{Introduction}
\label{sec:introduction}












Ethereum is a well-established, open-source blockchain platform that enables decentralized applications on the Internet, such as decentralized finance, voting systems, and non-fungible token marketplaces, through the creation and execution of smart contracts. 
These smart contracts are deployed onto the Ethereum blockchain network (i.e., being replicated across all participating nodes), facilitating autonomous and trustless execution of the contract's functions.
Smart contracts are immutable --- once deployed, they cannot be changed or tampered with. 
While immutability ensures the integrity and reliability of smart contract execution, it poses challenges for updating smart contracts (e.g., to fix bugs or introduce new features) since existing states (e.g., stored data, balances) must be migrated to new smart contracts.

\begin{table*}[t!]
\centering
\begin{tabular}{ |P{2.5cm}|P{1.5cm}|P{1.5cm}|P{1.5cm}| P{1.5cm}| P{1.5cm}  |P{1.5cm} | P{1.5cm}  |P{1.5cm}|}
    \cline{2-9}
    \multicolumn{1}{c|}{}&\multicolumn{4}{c|}{Smart contract coverage} & \multicolumn{4}{c|}{Collision coverage} \\
    \cline{2-9}
    \multicolumn{1}{c|}{}&\multicolumn{2}{c|}{With source code} & \multicolumn{2}{c|}{Without source code} & \multicolumn{2}{c|}{With source code} & \multicolumn{2}{c|}{Without source code} \\
    \cline{2-9}
   \multicolumn{1}{c|}{} & With tx & Without tx & With tx &  Without tx & Function & Storage & Function & Storage \\
    \cline{2-7}
    \hline
    \hline
    EtherScan~\cite{etherscan} & \cmark & \cmark &  &  &  &  & &\\
    Slither~\cite{feist2019slither} & \cmark & \cmark &  &  & \cmark & \cmark  & &\\
    Salehi et al.~\cite{salehi2022not} & \cmark &  & \cmark  &  & &  & &\\
    USCHunt~\cite{bodell2023proxy} & \cmark & \cmark &  &  & \cmark & \cmark & &  \\
    CRUSH~\cite{ruaro2024not} &  \cmark & &  \cmark  &  &  & \cmark & & \cmark \\
    Proxion (this work) & \cmark &  \cmark & \cmark  & \cellcolor{green!25} \cmark  & \cmark &  \cmark & \cellcolor{green!25} \cmark & \cmark\\
    \hline
 
 \hline
\end{tabular}
\caption{\codename uncovers more proxy smart contracts than previous works, especially including the {\em hidden contracts} that do not have source code and past transactions (acronym: tx) available. As a result, \codename also discovers more collision vulnerabilities, notably function collisions in contracts lacking source code. \codename's novel coverage is highlighted in green.}
\label{tab:coverage}
\end{table*}

To enable smart contracts' upgradeability while still adhering to their immutability, the {\em proxy design pattern} has recently emerged in several Ethereum Improvement Proposals (EIPs) (e.g.,~\cite{ethereum2023eip897,ethereum2023eip1167, ethereum2023eip1822, ethereum2023eip1967, ethereum2023eip2535}) as well as in major blockchain-based companies like OpenSea~\cite{opensea} and Compound~\cite{compound}.
Under this pattern, an original smart contract is decoupled into two contracts: a {\em proxy contract} that contains the data storage and a {\em logic contract} that contains the implementation logic.
The two contracts interact via delegate calls that allow the logic contract's functions to be executed in the context of the proxy contract's storage. 
To update the implementation of smart contracts under this scheme, developers simply deploy a new logic contract and change the logic contract's address stored in the proxy contract accordingly. 

The emerging proxy architecture also comes with new security issues, of which {\em function collisions} and {\em storage collisions} are the most notable ones. 
Particularly, smart contract developers may deliberately or accidentally create conflicts in the storage layouts or function identifiers between the proxy and logic contracts.
When users execute these colliding contracts, such conflicts can lead to stored data and functions being incorrectly accessed.
Worse, they can also be exploited by sophisticated adversaries to steal assets from the victims who wrongfully execute malicious functions or data.
For example, attackers can create malicious contracts with function collisions that disguise their scamming functionalities, often known as honeypot contracts~\cite{torres2019art}.
Outside of the academic realm, adversaries have leveraged storage collisions to overwrite the owner of Audius contracts, stealing more than a million worth of tokens in the process~\cite{audius}.
Also, a bounty hunter discovered storage collisions in the contract connecting the Ethereum and Arbitrum blockchains, which could potentially be exploited to compromise funds exceeding 250 million dollars~\cite{arbitrum}.

Considering the severe consequences of proxy smart contracts vulnerable to function and storage collisions in Ethereum, a few prior works have focused on finding them~\cite{etherscan,feist2019slither, salehi2022not, bodell2023proxy, ruaro2024not}.
Unfortunately, however, these research efforts fail to cover all smart contracts, thus missing out on potential collision exploits.
Notably, EtherScan~\cite{etherscan}, Slither~\cite{feist2019slither}, and USCHunt~\cite{bodell2023proxy} are limited to analyzing contracts with source code published by the developers.
Meanwhile, over 80\% of smart contracts are only available as deployed bytecode (i.e., the compiled source code), see Section~\ref{subsec:motivation}.
Moreover, CRUSH~\cite{ruaro2024not} and Salehi et al. \cite{salehi2022not} can identify proxy smart contracts only if they have previously interacted with logic smart contracts through transactions.
This constraint causes them to overlook nearly half of the active Ethereum contracts, particularly the newly deployed ones.

Failing to cover all smart contracts for proxy detection can lead to undesirable outcomes.
A clear consequence is that the scope of current tools in identifying collision problems is also limited.
Indeed, no prior research has successfully detected function collisions using only the bytecode of proxy and logic contracts.
For example, USCHunt~\cite{bodell2023proxy} and Slither~\cite{feist2019slither} are restricted to contracts with accessible source code, whereas CRUSH~\cite{ruaro2024not} is tailored to detect only storage collisions.
Worse, adversaries might deploy malicious contracts (e.g., honeypot contracts~\cite{torres2019art}) and hide them from these analysis tools by not publishing their source code or interacting with other contracts. 
These shortcomings highlight the necessity for a novel, effective assessment method for all contracts, thereby enabling thorough collision checks.


To that end, we introduce \codename, an automated cross-contract analyzer that aims to uncover all proxy smart contracts in Ethereum.
In essence, \codename emulates the execution of the contract under test with carefully crafted inputs that trigger distinct behaviors of proxy contracts (e.g., making delegate calls to other contracts).
Therefore, \codename does not require the contract's source code or its past transactions like prior works.
Once a proxy contract is identified, \codename efficiently locates the associated logic contracts throughout the blockchain history.
Subsequently, for each proxy and logic contract pair, \codename identifies function and storage collisions using various analysis techniques, depending on the availability of their source codes.
Table~\ref{tab:coverage} compares \codename with related works, highlighting its novel ability to cover hidden contracts lacking source code and historical transactions, as well as function collisions when contract source codes are unavailable.


In addition to providing broader coverage of smart contracts and collisions compared to previous studies, \codename stands out for its efficiency, effectiveness, and precision, as shown in our thorough evaluation. 
Specifically, \codename can analyze all 36 million active smart contracts in just 65 hours on a commodity server, processing an average of about 150 contracts per second. 
Moreover, \codename has identified thousands of vulnerable contracts with unreported collision issues, affecting at least 11 entities in control of a total of 19 billion USD in stakes. 
In terms of accuracy, \codename achieves 78.2\% in detecting storage collisions and 99.5\% in detecting function collisions, also surpassing the performance of state-of-the-art tools.


In summary, we claim the following contributions.

\begin{itemize}
\item We introduce \name, an automated cross-contract analyzer designed to assess proxy smart contracts in Ethereum for their function and storage collisions without the need for access to their source codes or historical transactions.

\item We implement \codename and demonstrate that it is efficient, effective, and accurate while covering significantly more contracts and collisions than state-of-the-art contract analyzers. 

\item 

We use \codename to analyze {\em all} alive Ethereum smart contracts, uncovering nearly 20 million proxy contracts, detecting 1.5 million collision issues, and capturing several insightful trends in developing proxy contracts over the years.

\end{itemize}
\section{Background}
\label{sec:background}

In this section, we provide the background by introducing the Ethereum blockchain (\S\ref{subsec:ethereum-blockchain}), reviewing proxy smart contracts (\S\ref{subsec:proxy-smart-contract}), and describing their collision issues (\S\ref{subsec:collision}).


\subsection{Ethereum Blockchain}
\label{subsec:ethereum-blockchain} 

Ethereum stands as a widely adopted cryptocurrency.
It operates on the principles of a distributed state machine where peer-to-peer nodes collaboratively maintain a global state consisting of users' accounts and balances.
A typical Ethereum node comprises a consensus client implementing the proof-of-stake algorithm and an execution client responsible for propagating and executing transactions within the Ethereum Virtual Machine (EVM), thereby deterministically generating a new blockchain state. 
These transactions serve the purpose of deploying specialized programs known as smart contracts onto the Ethereum network or executing functions within deployed smart contracts.

Specifically, a smart contract is a program written in high-level languages such as Solidity~\cite{solidity} or Viper~\cite{viper}, which is then compiled into EVM bytecode. 
Deploying a smart contract to the Ethereum blockchain essentially means creating a new contract account containing its compiled bytecode and data storage. 
The data storage stores variables consecutively based on the order in which they are declared.
Subsequently, users can trigger the EVMs to execute functions within the deployed smart contract by sending transactions that meet the predefined conditions, such as proper input data (i.e., call data\footnote{Call data is used to call a specific contract's function, which should not be confused with {\tt calldata}, which is the temporary location where function arguments are stored in the EVM.}), to the contract account's address.
These transactions are termed external transactions and can be initiated by any externally owned account that interacts with a contract on the blockchain. 
In contrast, internal transactions occur when a smart contract invokes the functions of another smart contract.

To execute a smart contract function, the transaction's call data encodes a series of bytes, including a function selector (also known as the function signature) and the function's arguments. 
The function selector is created by hashing the function's prototype string using the {\tt Keccak-256} hashing function and truncating the output to the first four bytes. 
For instance, the function selector for {\tt free\_ether\_withdrawal()} is {\tt 0xdf4a3106}. 
This selector is followed by a series of bytes representing the function's arguments, encoded depending on the compilers.

During the execution of smart contract bytecode, EVMs sequentially interpret and execute lower-level, machine-readable instructions known as opcodes. 
Specifically, the EVM extracts the function selector, parses its arguments, and executes the function body. 
If the call data's function selector does not match any defined function, the EVM will execute a fallback function if one is present in the smart contract. 
It is also worth noting that execution occurs within a stack machine, where a stack—acting as a list of 32-byte elements—holds the inputs and outputs of smart contract instructions.
Items added to the stack are placed at the top, and instructions typically interact with the stack's topmost elements. 
Additionally, during the execution, the EVMs may use a transient memory that is cleared after each transaction or access the persistent storage of the smart contracts.

\subsection{Proxy Smart Contracts in Ethereum}
\label{subsec:proxy-smart-contract}

Ethereum smart contracts are designed to be immutable, meaning their bytecode remains unchanged after being deployed on the blockchain. 
Nonetheless, similar to conventional programs, smart contracts require updates to introduce new features, correct errors, or address security flaws.
A naive solution to updating a smart contract's logic is migrating its states and balances into a new contract with the updated program. 
Obviously, such a migration does not scale, as all affected users need to update their workflows to use a new contract address. 
The need for smart contract upgradability motivates several proposals for splitting a smart contract into two parts: a {\em proxy} contract storing the data and a {\em logic} contract storing the implementation logic~\cite{ethereum2023eip897, ethereum2023eip1167, ethereum2023eip1822, ethereum2023eip1967, ethereum2023eip2535}.

\begin{figure}[t!]
\begin{center}
  \includegraphics[width=0.48\textwidth]{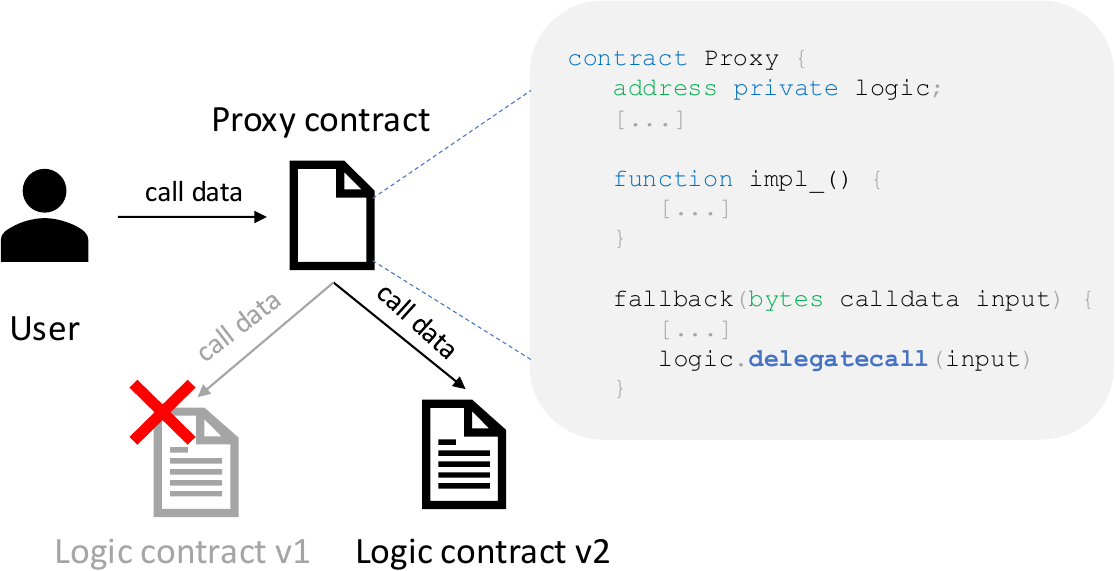}
  \caption{An example of proxy and logic smart contracts. The proxy contract's delegate call forwards the call data to the upgraded logic contract.
  }
  \label{subfig:proxy-smart-contract}
\end{center}
\end{figure}

We illustrate the essence of proxy smart contracts in Figure~\ref{subfig:proxy-smart-contract}.
The proxy smart contract facilitates a delegate call to a function in the logic contract.
Here, the delegate call allows the execution of the logic contract's function in the context of the proxy contract's storage. 
In particular, the user encodes the call data for that function in an Ethereum transaction sent to the proxy contract. 
This call data contains a function selector that does not match any existing proxy contract functions, allowing it to be passed to the proxy contract's fallback function. 
As a result, the delegate call in the fallback function is triggered, executing the logic contract's functions while accessing the proxy contract's storage. 

Naturally, an important application of proxy smart contracts is upgrading the logic implementations.  
To enable upgradeability, the proxy contract stores the logic contract's address, usually in one of its storage slots, see {\tt logic} variable in Figure~\ref{subfig:proxy-smart-contract}.
When upgrading to a new logic contract (e.g., v2 in Figure~\ref{subfig:proxy-smart-contract}), the user only needs to replace the logic contract's address stored in the proxy contract (e.g., using a {\tt setter} call).
Another notable application of proxy smart contracts is cloning exact contract functionalities~\cite{ethereum2023eip1167}.
Often, these proxy contracts contain no function but only delegate calls to a fixed logic contract's address stored directly in the bytecode.

This paper excludes library contracts with reusable code for general use, such as SafeMath~\cite{safemath}, from the definition of logic smart contracts and, therefore, does not automatically classify contracts that call them as proxy smart contracts.
Specifically, in these external library calls, delegate calls occur at different code locations, not in the fallback function. 
This eliminates the possibility of upgradability, as it prevents users from choosing whether to execute the library contract's functions. 
Indeed, all Ethereum Improvement Proposals (EIPs) concerning the proxy smart contract pattern exclude library contracts from their scope.

In short, \codename considers a contract to be a proxy contract if it uses the delegate call in its fallback function to forward the call data it has received to another contract, and any contract receiving forwarded call data from a proxy contract is a logic contract.

\subsection{Collision Vulnerabilities}
\label{subsec:collision}


Separating storage and implementation logic in proxy architecture leads to various issues, notably function and storage collisions. 
These collisions have posed significant risks to proxy and logic contracts, potentially enabling attacks that aim to steal stored assets.

\begin{figure}[t!]
\begin{center}

\begin{lstlisting}[language=Solidity, caption={An example of function collisions between the proxy and logic contracts. The proxy contract's {\tt impl\_LUsXCWD2AKCc()} function collides with a logic contract's {\tt free\_ether\_withdrawal()} due to their same signatures.},label={fig:function-collision}]
contract Proxy {
    address private owner;
    address private logic; // Logic contract's address
    address constant USDT = 
        0xdAC17F958D2ee523a2206206994597C13D831ec7;
    
    constructor(address impl) {
        owner = msg.sender;
        logic = impl;
    }
    
    function impl_LUsXCWD2AKCc() public {
        // a malicious function stealing 1000 USDT from the caller
        USDT.delegatecall(abi.encodeWithSignature(
            "transfer(address,uint256)", owner, 1000
        ));
    }
    
    fallback(bytes calldata input) external
             returns (bytes memory) {
        (bool success, bytes memory output) = 
            logic.delegatecall(input);
        return output;
    }
}

contract Logic {
    function free_ether_withdrawal() public {
        // an attractive function that sends the caller 10 Ethers
        payable(msg.sender).transfer(10 ether);
    }
}

\end{lstlisting}
\end{center}
\end{figure}

\para{Function collision vulnerability}
A {\em function collision} occurs when a proxy contract's function has the same signature as a logic contract's function.
When it happens, users cannot execute the collided functions in the logic contract because the call data is not passed to the fallback function in the proxy contract.
The most obvious scenario of function collisions is when a proxy contract's function has the same name as a logic contract's function.
The two collided functions may also differ as long as they share the same first 4 bytes in their hashes.
Listing~\ref{fig:function-collision} illustrates an example of function collisions, in which two functions {\tt impl\_LUsXCWD2AKCc()} (line 12) and {\tt free\_ether\_withdrawal()} (line 28) have the same 4-byte signature of {\tt 0xdf4a3106}.
As a result, when the user encodes this signature in the call data, the proxy contract will execute {\tt impl\_LUsXCWD2AKCc()} instead of calling the logic contract's {\tt free\_ether\_withdrawal()}.

{\em Potential exploits.}
Malicious contract developers can exploit function collisions to trick users into executing honeypot contracts~\cite{torres2019art}. 
In such attacks, the adversary creates a logic contract with an enticing function, such as transferring cryptocurrencies to the function caller, which actually collides with a proxy contract's function that steals funds from the caller.
We exemplify the honeypot contracts in Listing~\ref{fig:function-collision}, where function {\tt free\_ether\_withdrawal()} in the logic contract allows the caller to withdraw 10 ETH from the contract's balance (line 30).
However, since this function has the same function selector with  {\tt impl\_LUsXCWD2AKCc()}, the user transfers 1,000 USDT to the contract owner instead (lines 14--16).

We note that creating a pair of functions that share the same 4-byte signature is remarkably easy and achievable within seconds on even modest computers.
Attackers might also craft functions that follow a certain naming pattern or share the same signature with existing functions, requiring more time. 
For instance, after approximately 600 million attempts in 1.5 hours with a commodity laptop, we found a function with the same signature as {\tt free\_ether\_withdrawal()}.
Overall, the process of finding colliding functions remains highly accessible for motivated adversaries, emphasizing the significant risks posed by function collisions.

\para{Storage collision vulnerability}
A {\em storage collision} happens when two variables with different types or interpretations are assigned to the same storage slots across proxy and logic contracts. 
To see why such collisions may occur, recall that the logic contract is executed in the context of the proxy contract's storage, meaning that two contracts share the same storage layout.
For example, the first variable declared in the logic contract, regardless of its name and type, will access the storage slot 0 of the proxy contract.
Commonly, if multiple contiguous variables require less than 32 bytes (for example, a {\tt bool} is 1 byte or an {\tt address} is 20 bytes), they will be packed into a single storage slot (e.g., in Solidity).
Thus, collisions often come from mismatched storage layouts (i.e., the orders of variable declarations) and can cause data to be incorrectly read or overwritten. 
The most frequent cause of storage collisions is when one contract writes to a slot, and another reads from that slot with a different interpretation.
Upgrading the logic contract to newer versions that change the order or types of variables also facilitates storage collisions.
We show an example of storage collisions in Listing~\ref{fig:storage-collision} where proxy contract's {\tt owner} variable (20 bytes) and logic contract's {\tt initialized} and {\tt initializing} variables (1 byte each) use the same storage slot $0$.
Thus, when the user executes a logic contract function involving the {\tt initialized} variable, it may access 1 byte of the {\tt owner} variable in the proxy contract.

\begin{figure}[t!]
\begin{center}

  \begin{lstlisting}[language=Solidity, caption={An example of storage collisions between the proxy and logic contracts. The storage collision occurs at slot $0$ between the {\tt owner} variable (20 bytes) in the proxy contract versus {\tt initialized} and {\tt initializing} variables (1 byte each) in the logic contract.},label={fig:storage-collision}]
contract Proxy {
    address private owner; // Storage slot [0x0]
    [...]
    address private logic; // Logic contract's address

    constructor(address impl) {
        logic = impl;
    }
    [...]
    fallback(bytes calldata input) external
             returns (bytes memory) {
        (bool success, bytes memory output) = 
            logic.delegatecall(input);
        return output;
    }
}

contract Logic {
    bool private initialized; // Storage slot [0x0]
    bool private initializing; // Storage slot [0x0]

    function initialize() external {
        require(initializing || !initialized);
        initialized = true;
        initializing = false;
        owner = msg.sender;
    }
    [...]
}

\end{lstlisting}
\end{center}
\end{figure}

{\em Potential exploits.} 
Storage collisions can be exploited to seize control of vulnerable contracts by overwriting the owner's address with that of the attacker~\cite{audius}. 
Additionally, adversaries can deceive users into executing malicious logic contracts in which variables may have names that seem harmless but are designed to access storage slots in the proxy contract, leading to actions that differ from the user's expectations.

For example, Listing~\ref{fig:storage-collision} presents a simplified illustration of the vulnerable proxy and logic contracts exploited in the real-world attacks against the Audius cryptocurrency~\cite{audius}. 
In this example, the logic smart contract contains an {\tt initialize()} function that sets the transaction's sender as the owner of the contract (line 26) if the owner has not been set previously (line 23).
This function is intended to be called only once during contract deployment.
However, the {\tt owner} variable in the proxy contract (line 2) and the {\tt initialized} and {\tt initializing} variables in the logic contract are both allocated to the same storage slot number 0.
Consequently, even after the {\tt initialized} and {\tt initializing} variables are updated (lines 24--25), indicating the owner has been assigned, the storage slot is immediately overwritten by the new {\tt owner} value (line 26) in the proxy contract.
Consequently, the {\tt initializing} variable in the logic contract is always {\tt true}, wrongly indicating that the contracts have not completed the initialization.
This allows the {\tt initialize()} function to be executed multiple times and the {\tt owner} variable to be reassigned.
Attackers indeed exploited this vulnerability to take control of the Audius governance contracts, as detailed in Audius's post-mortem report~\cite{audius}.
\section{\codename Overview}
\label{sec:overview}

This section provides an overview of our tool \codename. 
First, we discuss the necessity for a novel tool to uncover proxy smart contracts within Ethereum and outline the challenges that current solutions fail to address (\S\ref{subsec:motivation}). 
We then briefly present the methods by which \codename overcomes these challenges (\S\ref{subsec:solutions}).

\subsection{Motivation}
\label{subsec:motivation}

Proxy smart contracts are crucial in the Ethereum ecosystem due to their upgradability, yet they are vulnerable to exploitation if collision vulnerabilities are present. 
Unfortunately, the best practices of proxy smart contracts fall short for both developers and users.
Specifically, users must manually review the source code of both proxy and logic contracts before any interaction, such as sending a transaction. 
Should a proxy contract lack available source code, it is advised that users avoid engaging with it to prevent potential misuse despite possibly missing out on legitimate services.
Developers of proxy smart contracts are recently equipped with the new Transparent Upgradeable Proxy design proposed by OpenZeppelin~\cite{openzeppelin} that minimizes the impacts of function collisions.\footnote{In the Transparent Upgradeable Proxy pattern, the owner of the contracts is able to execute all functions with the exception of the fallback function, whereas other users always delegate calls. 
This architecture, by distinguishing between the callers, prevents unintended function executions and inherently avoids function collisions.}
However, adopting a new proxy design is time-consuming, while existing proxy contracts may already contain vulnerabilities due to human errors.

To facilitate the safety assessment of proxy smart contracts, various systems have been recently introduced to automatically identify collision vulnerabilities~\cite{feist2019slither, bodell2023proxy, ruaro2024not}.
At the high level, these systems typically involve two main phases: (1) identifying proxy contracts and their corresponding logic contracts from historical data, and (2) examining each contract pair to determine if there are any colliding functions or storage slots. 
While the approach seems straightforward, several challenges remain, leaving a non-negligible number of proxy smart contracts unchecked and potentially vulnerable to exploitation.




\begin{figure}[t!]
\begin{center}
  \includegraphics[width=0.48\textwidth]{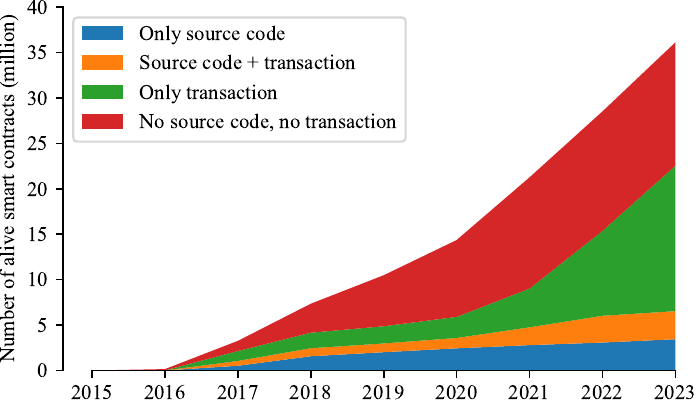}
  \caption{The accumulated number of alive Ethereum smart contracts till 31 October 2023.
  Prior works only cover about 18\% of smart contracts with source code (blue and orange)~\cite{bodell2023proxy} or 53\% of smart contracts with past transactions (orange and green)~\cite{ruaro2024not}, while \codename also applies to the hidden contracts without source code or past transactions (red).
  }
  \label{fig:smart-contract-ratios}
\end{center}
\end{figure}

First, the source code of smart contracts may not be available because the public blockchain contains only their runtime bytecode.
The contract bytecode is not human-readable, making the analysis (e.g., checking for delegate calls in the fallback function) difficult.
Indeed, several existing tools (e.g., USCHunt~\cite{bodell2023proxy}, Slither~\cite{feist2019slither}) can only analyze smart contracts when their source code is published (e.g., on EtherScan~\cite{etherscan}).
Figure~\ref{fig:smart-contract-ratios} shows the accumulated number of {\em alive}\footnote{We exclude the destroyed smart contracts.} Ethereum smart contracts from 2015 to 2023.
Unfortunately, we notice that smart contracts with source code available only account for less than 20\% of all contracts.

Second, uncovering proxy smart contracts through their historical interactions with other (logic) smart contracts might also not always be feasible.  
Specifically, existing tools (e.g., CRUSH~\cite{ruaro2024not}, Salehi et al.~\cite{salehi2022not}) analyze all blockchain transactions to detect \delegate instructions, identifying contracts involved as proxy and logic contract pairs.
This approach is applicable to all smart contracts, even those with only bytecode available. 
However, it can result in numerous false positives, as common library calls may also contain \delegate instructions.
Furthermore, many smart contracts have never interacted with others, such as those freshly deployed on the blockchain. 
In fact, our data in Figure~\ref{fig:smart-contract-ratios} indicates that only about half of the active smart contracts have had interactions with other contracts and are therefore detectable by these tools.
Intuitively, adversaries may exploit these by concealing malicious proxy contracts, such as not releasing their source code and avoiding prior interactions with other smart contracts.

Third, even when proxy and logic smart contracts with only bytecode are uncovered (e.g., using CRUSH~\cite{ruaro2024not}), detecting their function collisions is highly error-prone due to the lack of information obtained from their bytecode.
As we will show shortly below, function signatures, which are four bytes long, always appear after a specific byte (i.e., the {\tt PUSH4} opcode) in the contract's bytecode. 
A naive function collision detection would cross-check any four bytes following the {\tt PUSH4} opcode in the proxy and logic contracts pair.
However, this method is flawed as arbitrary data may also follow the {\tt PUSH4} opcode, leading to numerous false positives.
For this reason, no previous research has managed to detect function collisions solely from bytecode; for example, USCHunt~\cite{bodell2023proxy} and Slither~\cite{feist2019slither} are limited to contracts with available source code, whereas CRUSH~\cite{ruaro2024not}  is designed to identify storage collisions only.

\subsection{Solutions}
\label{subsec:solutions}



In this paper, we propose \codename, an automation tool that aims to reveal all proxy smart contracts and their corresponding logic contracts.
The main novelty of \codename is its capability to identify the {\em hidden proxy smart contracts} that lack both source code and previous transactions.
To achieve that, \codename employs dynamic analysis to verify whether the delegate calls forward the transaction call data in the fallback function.
Specifically, for a given smart contract, \codename emulates its EVM execution using carefully crafted call data.
If a smart contract is indeed a proxy, a \delegate instruction will appear in the EVM stack and vice versa.
Through the emulation of EVM execution, \codename can also identify the storage locations of the logic contracts' addresses, enabling their easy retrieval from historical blockchain data.
Consequently, unlike previous methods, \codename is not dependent on the source code or historical transactions of the smart contracts under test.

Thereafter, given a pair of proxy and logic smart contracts, \codename searches for potential storage and function collisions.
For storage collisions, \codename utilizes symbolic execution and program slicing techniques from the state-of-the-art tool CRUSH~\cite{ruaro2024not} to identify exploitable contracts.
Regarding function collisions, \codename compares the function signatures pairwise using Slither~\cite{feist2019slither} if both proxy and logic contracts have their source code available.
Here, another innovation of \codename lies in detecting function collisions, even when one or both of the contracts do not have available source code.
Specifically, when a contract exists solely in bytecode, \codename examines the disassembled opcodes to identify the jump instructions corresponding to code blocks of functions. 
\codename then extracts the 4-byte data of the function signature that precedes these jump instructions.
While the exact function names remain undisclosed, retrieving these signatures is sufficient for \codename to cross-reference and detect any function collision.


A prototype of \codename, featuring the proxy smart contract finder and the collision detector, is accessible at {\em \url{https://github.com/Proxion-anonymous/Proxion}}.

\section{Uncovering Proxy Smart Contracts}
\label{sec:uncovering-psc}

\begin{figure}[t!]
\begin{center}
  \includegraphics[width=0.48\textwidth]{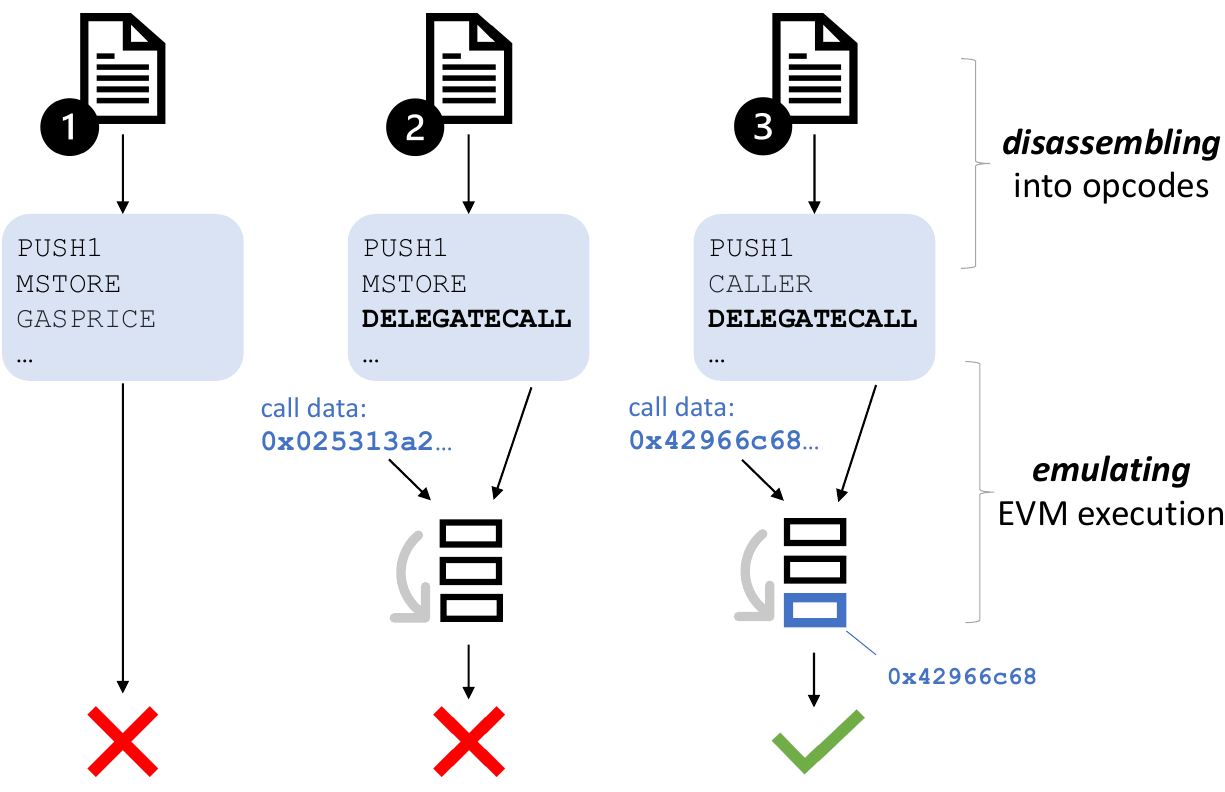}
  \caption{\codename identifies proxy smart contracts in two steps. 
  First, \codename disassembles the tested contract's bytecode into opcodes. 
  Contracts without a {\tt DELEGATECALL} opcode (e.g., \ding{202}) are not proxies. 
  Second, \codename executes the tested contract under an emulated EVM with carefully created transaction call data.
  If this data is not forwarded to another contract in the emulated EVM, the contract is not a proxy (e.g., \ding{203}) and vice versa (e.g., \ding{204}).
  \codename then identifies the associated logic contracts of the identified proxy contract.
  }
  \label{fig:identifying-proxy}
\end{center}
\end{figure}

We illustrate the two-step approach to check if a given smart contract is a proxy contract in Figure~\ref{fig:identifying-proxy}.
The approach includes disassembling smart contracts to locate \delegate instructions (\S\ref{subsec:deassembler}) and emulating the EVM execution to verify the contract's interaction with other contracts (\S\ref{subsec:evm-emulator}).
Then, \codename also identifies all logic contracts associated with identified proxy contracts from historical blockchain data (\S\ref{subsec:finding-logic-contracts}).

\subsection{Disassembling Smart Contracts}
\label{subsec:deassembler}

\begin{figure}[t!]
\begin{center}    
\begin{lstlisting}[language=Solidity, caption={The disassembled opcodes for the proxy smart contract's bytecode presented in Listing~\ref{fig:function-collision}.},label={fig:opcode},breaklines=true]
0000  60  PUSH1 0x80
0002  60  PUSH1 0x40
0004  52  MSTORE
0005  34  CALLVALUE
0006  80  DUP1
0007  15  ISZERO
0008  61  PUSH2 0x000f
000B  57  *JUMPI
[...]
// function selection process
001A  35  CALLDATALOAD
001B  60  PUSH1 0xe0
001D  1C  SHR
001E  80  DUP1
001F  63  PUSH4 0xdf4a3106 // signature of impl_LUsXCWD2AKCc()
0024  14  EQ
0025  61  PUSH2 0x00ce
0028  57  *JUMPI
[...]
// fallback():
007C  5B  JUMPDEST
007D  5F  5F
007E  60  PUSH1 0x40
0080  51  MLOAD
0081  80  DUP1
0082  83  DUP4
0083  03  SUB
0084  81  DUP2
0085  85  DUP6
0086  5A  GAS
0087  F4  DELEGATECALL
[...]
// impl_LUsXCWD2AKCc():
00CE  5B  JUMPDEST
[...]

\end{lstlisting}
\end{center}
\end{figure}


In the first step, \codename determines the tested smart contract is not a proxy if its bytecode does not contain the \delegate opcode, which is the defining factor of all proxy smart contracts. 
To learn the opcodes of a smart contract, \codename disassembles its bytecode, which results in a sequence of assembly representation known as opcodes and operands (e.g.,~\cite{grech2019gigahorse, grech2022elipmoc}).

We implement this disassembler component of \codename based on Octopus, an open-source security analysis framework that is capable of translating contract bytecode into opcodes and operands~\cite{octopus}. 
Particularly, we extend Octopus so that it covers recently introduced opcodes in Ethereum, such as {\tt CALL}, {\tt DELEGATECALL}, {\tt CREATE}, and {\tt CREATE2}.
This is easily achievable since the opcodes have fixed corresponding bytes.
Listing~\ref{fig:opcode} illustrates an example of the opcodes after being disassembled from the proxy contract's bytecode shown previously in Listing~\ref{fig:function-collision}.

Thereafter, \codename spots if any \delegate opcode exists, concluding the smart contract is not a proxy (e.g., contract \ding{202}) or proceeding to the next step (e.g., contracts \ding{203} and \ding{204}).

\subsection{Emulating EVM Execution}
\label{subsec:evm-emulator}

In the second phase, \codename checks if delegate calls are triggered in the tested smart contract's fallback function, and they indeed forward the transaction call data to another smart contract. 
To do so, \codename triggers the tested smart contract with an emulated EVM and carefully crafted call data. 
Particularly, this call data contains a random function signature (i.e., with 4 bytes in size) that is different from signatures of all other functions in the proxy contract.
Thus, it enables the execution of the proxy contract's fallback function. 
To learn the potentially existing functions' signatures, \codename identifies the locations of {\tt PUSH4} opcodes in the contract's bytecode and extracts the 4-byte data following each of them.
This approach is based on an observation that popular contract compilers (e.g., Solidity, Vyper) always include the function signatures following {\tt PUSH4} opcodes.
While not all 4-byte data following {\tt PUSH4} opcodes is a function signature, \codename safely avoids all of them. 
Next, \codename emulates the EVM execution of the tested smart contract along with the generated transaction call data and observes the memory, stack, and storage of each instruction. 
If \codename does not observe this data is passed to the logic contract's context after the execution of the \delegate instruction, \codename marks the tested smart contract as not a proxy (e.g., contract \ding{203}).
Otherwise, the tested smart contract is a proxy (e.g., contract \ding{204}).

Here, we again extend Octopus to implement our EVM emulator. 
Specifically, our EVM emulator can also handle opcodes that have values depending on the state of the blockchain.
For example, to support the {\tt NUMBER} opcode that pushes the current block's number to the EVM stack, we use the values from the latest block on the blockchain since all alive contracts are supposed to be executable at any block's numbers.
Similarly, we use the values in the latest block for the {\tt BLOCKHASH}, {\tt DIFFICULTY}, {\tt GASLIMIT}, {\tt TIMESTAMP}, and {\tt GASPRICE} opcodes. 
We also assign fixed values for a few other opcodes, such as {\tt CHAINID}, {\tt BASEFEE}, and {\tt COINBASE}, using the most probable values (e.g., the chain ID of Ethereum's mainnet is 1).
Adding these blockchain-related opcodes enhances the fidelity of the EVM emulation (e.g., with fewer runtime errors when encountering them).

Moreover, we implement our EVM emulator to support {\tt CALL} and \delegate opcodes that specifically call another contract and obtain the execution results before proceeding. 
To do so, we create two EVM emulator instances, one for the caller and another for the callee, and copy the results back from the callee to the stack of the caller instance to simulate the function returning.

For the opcodes that place bytecode on Ethereum at a smart-contract address (e.g., \texttt{CREATE} and \texttt{CREATE2}), we use a fixed address to ensure that we can retrieve the exact address of a newly created contract.
If our EVM emulator encounters this fixed address, we treat it like a normal smart contract. 
This method is acceptable because of the negligible probability of address collision (e.g., only 1 out of $2^{160}$ in Ethereum).

\subsection{Finding associated logic contracts}
\label{subsec:finding-logic-contracts}

Upon identifying a proxy smart contract, \codename finds its associated logic contracts, which also can be done by looking into the EVM stack when the \delegate instruction is executed.
Indeed, the address of the current logic contract is one of the stack inputs following the \delegate instruction.

Next, \codename also finds all other logic smart contracts that are ever associated with the tested proxy contract.
If \codename observes the found logic contract's address is hard-coded in the proxy contract's bytecode, it considers the test proxy contract follows the {\em minimal proxy pattern} (i.e., EIP-1167).
These minimal proxy contracts include no function but only a delegate call in the fallback function and fix the address of the logic contract in the bytecode. 
Thus, they are lightweight (e.g., their bytecode is less than 100 bytes) and have only one associated logic smart contract throughout history.

\begin{algorithm}[t!]
  \caption{Finding addresses contained in a storage slot.}
  \label{alg:finding-logic-contracts}
  \begin{algorithmic}[1]
    \small
    \Require{
    $\mathcal{PSC}$: The tested proxy smart contract. \newline
    $h_{lower}, h_{upper}$: The lower and upper bounds for the height of considered blocks (e.g., the genesis block and the latest block).
    }
    \Ensure{$\mathcal{A}$:  The set of logic contracts' addresses associated with $\mathcal{PSC}$.}
    \newline
    \Procedure{PartitionBlocks($h_{lower}, h_{upper})$}{}
        \Let{\mathcal{V}_{lower}}{\mathsf{getStorageAt}(\mathcal{PSC}, h_{lower})}
        \Let{\mathcal{V}_{upper}}{\mathsf{getStorageAt}(\mathcal{PSC}, h_{upper})}
        
        \If{$\mathcal{V}_{lower}=\mathcal{V}_{upper}$}\Comment{Storage slot values are the same.}
          \State\Return{$\{V_{lower}\}$}
        \EndIf

        \Let{h_{mid}}{\lfloor(h_{lower}+h_{upper})/2\rfloor}\Comment{Binary search.}

        \Let{\mathcal{A}_{lower}}{\textsc{PartitionBlocks}(h_{lower}, h_{mid})}
        \Let{\mathcal{A}_{upper}}{\textsc{PartitionBlocks}(h_{mid}+1, h_{upper})}

        \Let{\mathcal{A}}{\mathcal{A}_{lower} + \mathcal{A}_{upper}- \{\varnothing\}}
        \State\Return{$\mathcal{A}$}

    \EndProcedure
  \end{algorithmic}
\end{algorithm}

If the found logic contract's address is in a proxy contract's storage slot, \codename employs a binary search to uncover all addresses that have been stored in that slot. 
Naively, one can utilize Ethereum APIs such as {\tt getStorageAt} to check the slot's content at specific blockchain height (e.g., from the genesis block to the latest one), which is time-consuming when millions of Ethereum blocks exist. 
Instead, \codename assumes that logic contracts of the same proxy contract are unique throughout history since reusing old versions of logic contracts (e.g., containing bugs or missing features) is intuitively uncommon. 
Leveraging this observation, \codename implements a binary search for blocks in which the value of the storage slot changes.

We illustrate how \codename finds the logic contracts' addresses in Algorithm~\ref{alg:finding-logic-contracts}.
Particularly, \codename starts with the genesis block and the latest block and compares the storage slot values at these two blocks (i.e., lines 2--3).
If the values are the same, the storage slot does not change during this block range (i.e., lines 4--6).
Otherwise, \codename splits the range into two halves and repeats this procedure (i.e., lines 7--9). 
\codename finally returns a set of all values ever stored in the tested proxy smart contract; see line 11. 

\section{Checking For Collisions}
\label{sec:checking-collision}

Once \codename identifies proxy and logic contracts, it detects if they are susceptible to function (\S\ref{subsec:function-collision}) and storage collisions (\S\ref{subsec:storage-collision}).

\subsection{Function collisions}
\label{subsec:function-collision}

Once the proxy contract and its associated logic contracts are identified, \name utilizes Etherscan~\cite{etherscan}, a widely-used Ethereum explorer, to obtain their source code.
To maintain a uniform format for the contract source code, we have developed a parser that processes the source code provided by the Etherscan APIs, which may be in the form of dictionaries or arrays.

For a pair of proxy and logic contracts with available source code, \codename implements a static analyzer based on Slither~\cite{feist2019slither} to detect their function collisions. 
In particular, we generate a list of signatures for all functions in each contract.
If there is an intersection between these lists, the contracts are considered to have a collision.
To speed up the collision detection in a large dataset of contracts (see Section~\ref{sec:landscape}), \codename also groups contracts based on their bytecode hash, indicating that the contracts are identical despite being deployed at different addresses.

When one or both of the proxy or logic contracts do not have source code, \codename analyzes the disassembled opcodes instead.
It is important to remember that function signatures are always preceded by a {\tt PUSH4} opcode, although the converse is not true (i.e., the data following a {\tt PUSH4} opcode can be arbitrary). 
Therefore, the challenge lies in identifying which 4-byte data subsequent to the {\tt PUSH4} opcodes actually constitutes a function signature.
To achieve this, \codename begins by identifying the jump instructions (e.g., {\tt JUMP} or {\tt JUMPI} opcodes), which divide the disassembled code into several basic blocks.
These code blocks may represent if-else statements, loops, or function calls, which are distinguishable by how the EVM execution reaches them via the jump instructions.
In particular, the EVM execution typically jumps to a specific function after a condition involving its function signature is satisfied (e.g., when call data contains that signature). 
Thus, \codename searches for a pattern of opcodes containing {\tt PUSH4} (i.e., pushing 4 bytes), {\tt EQ} (i.e., equal), and {\tt *JUMPI} (i.e., conditional jump), see line 15--18 in Listing~\ref{fig:opcode}.
\codename then extracts the 4-byte sequence following the {\tt PUSH4} opcode within these patterns, treating it as the function signature.
Once \codename has gathered function signatures from both the proxy and logic contracts, \codename simply cross-checks them pairwise, similarly to when both contracts have source code.


In terms of implementation, \codename utilizes the state-of-the-art decompiler tool Panoramix~\cite{panoramix}, an integral part of Etherscan,  to disassemble the bytecode and identify the code blocks. 
\codename then parses Panoramix's outputs to identify the functions and retrieve their signatures.








\subsection{Storage collisions}
\label{subsec:storage-collision}

\codename utilizes techniques from CRUSH~\cite{ruaro2024not} to identify exploitable storage collisions.
To ensure a comprehensive understanding, we briefly explain the fundamental operations of CRUSH.

First, CRUSH identifies the storage slots possibly accessed by the {\tt SLOAD} or {\tt SSTORE} instructions, which can be found after disassembling the contract's bytecode. 
Subsequently, CRUSH employs program slicing to extract all instructions that contribute to the computation of the storage slots.
Following this, CRUSH symbolically executes these instructions to learn the size of the variables within the specified slots, thereby deducing their types.
Then, CRUSH compares all storage slots pairwise to uncover any type discrepancies suggesting a possible collision.

Upon detecting storage collisions, CRUSH identifies those that are potentially exploitable. 
CRUSH specifically targets sensitive storage slots involved in access control decisions, such as read-only slots or those written exclusively by certain users. 
It then generates test transactions to trigger collisions in these sensitive slots by writing one variable type and then reading a different type from the same slot. 
These transactions are subsequently fed to the EVM execution to verify the exploit.
\section{Evaluation}
\label{sec:evaluation}

This section evaluates \codename's performance, effectiveness, and accuracy.
We first demonstrate that \codename is efficient, capable of performing hundreds of proxy contract checks per second on a commodity computer (\S\ref{subsec:performance}). 
Next, we show that \codename effectively identifies unknown proxy and logic contracts that are susceptible to function and storage collisions within already examined datasets (\S\ref{subsec:effectiveness}).
Lastly, we demonstrate that \codename's accuracy surpasses that of the state-of-the-art tools (\S\ref{subsec:accuracy}).

\subsection{Performance}
\label{subsec:performance}

\para{Setup}
We operate \codename on a system equipped with Ubuntu 22.04 OS, featuring 12 cores (24 threads) at 3.8 GHz each, and 64 GB of RAM.
The input for \codename is a list of all 36 million active contract addresses in Ethereum (see Section~\ref{sec:landscape}).
Its output includes a list of proxy contracts and the logic contracts associated with each of them.
Additionally, \codename provides details of function and storage collisions for each proxy and logic contract pair, if such collisions exist.

\para{Results}
On average, \codename analyzes a smart contract in just 6.4 milliseconds to determine if it is a proxy, translating to 156.3 contracts per second. 
This efficiency enables \codename to process roughly 36 million active contracts on the Ethereum network within approximately 65 hours.

Moreover, \codename's binary search method for identifying logical contracts linked to a proxy contract significantly increases its efficiency. 
For each proxy contract, \codename makes 26 {\tt getStorageAt} API calls on average, which is a substantial improvement over the naive method of querying all 15 million Ethereum blocks.

In terms of function collision detection, \codename averages 6.7 milliseconds to check if a contract pair has a collision issue. 
The storage collision detection based on CRUSH averages a longer execution time of 1.3 minutes per contract pair. 
\codename overcomes this performance bottleneck by leveraging previously analyzed contract information with the same bytecode.
Indeed, by avoiding re-testing identical contracts, \codename requires only 48 days to test the storage collision vulnerabilities of 36 million smart contracts.




\subsection{Effectiveness}
\label{subsec:effectiveness}

\para{Datasets}
We evaluate the effectiveness of \codename in uncovering more proxy and logic contracts (and subsequently, more collisions) by comparing it with state-of-the-art tools, specifically USCHunt~\cite{bodell2023proxy} and CRUSH~\cite{ruaro2024not}, utilizing their original datasets.

First, we use the Smart Contract Sanctuary dataset~\cite{smart_contract_sanctuary}, which is used to evaluate USCHunt originally.
The Smart Contract Sanctuary dataset contains 329,764 smart contracts that were deployed between 2017 and 2022 and have their source code published on EtherScan~\cite{etherscan}.

Second, we also use the dataset used by CRUSH~\cite{ruaro2024not} that includes 53,580,899 contracts deployed from July 2015 to April 2023. 
These contracts may or may not have source code and past transactions. 

\para{Results}
\codename discovers more proxy contracts than both USCHunt and CRUSH when testing against their respective datasets.

When running with the entire Smart Contract Sanctuary dataset, we observe that \codename experiences notably fewer failure cases than USCHunt.
Specifically, USCHunt encounters halt due to contract compilation errors (e.g., unknown compiler versions) in about 30\% of cases.\footnote{We run USCHunt with the default compiler flags. There may be fewer errors if the compiler versions are provided when compiling each contract.}
Meanwhile, \codename fails to emulate the execution, for instance, due to insufficient values on the EVM stack in only about 1.2\% of contracts.
In total, \codename identifies 35,924 proxy contracts, whereas USCHunt detects only 29,023, which is roughly seven thousand fewer.
{\em As a result, \codename detects 257 function collisions that USCHunt has not reported}.

Furthermore, CRUSH indicates that 26.6\% of the examined smart contracts, totaling 14,237,696, are identified as proxy contracts.
Among them, CRUSH also detects that 956 contract pairs are vulnerable to storage collisions with verified exploits.
\codename reports about 1.2 million fewer proxy contracts than CRUSH in this dataset, totaling 13,042,496 proxy smart contracts.
This outcome occurs because CRUSH categorizes any contracts that involve \delegate instructions as proxy contracts, including the one making library calls.
In contrast, \codename does not consider this condition when classifying proxy contracts (cf. Section~\ref{subsec:proxy-smart-contract}).
When excluding those proxy smart contracts, \codename uncovers more 1,667,905 proxy contracts than CRUSH does, none of which have past transactions available.
{\em Moreover, \codename identifies an additional 1,480 contracts with exploitable storage collisions that CRUSH did not report.}

\subsection{Accuracy}
\label{subsec:accuracy}

\begin{table}[t!]
\centering
\normalsize
\begin{tabular}{|c|c|c|c|c|c|c|}
  \cline{3-7}
  \multicolumn{2}{c|}{} & \textbf{TP} & \textbf{FP} & \textbf{TN} & \textbf{FN} & \textbf{Accuracy}\\
  \cline{3-7}
  \hline
  \hline
  \multirow{3}{4em}{\bf Storage collision} &USCHunt & 33 & 83 & 79 & 11 & 54.4\% \\
  \cline{2-7}
  &CRUSH & 26 & 76 & 86 & 18 & 54.4\%   \\
  \cline{2-7}
  &\codename & 27 & 28 & 134 & 17 & \cellcolor{green!25}78.2\%\\
  \hline
  \hline
  \multirow{2}{4em}{\bf Function collision} &USCHunt & 299 & 1 & 0 & 261 & 53.3\% \\
  \cline{2-7}
  &\codename & 557 & 0 & 1 & 3 & \cellcolor{green!25}99.5\%\\
  \hline
\end{tabular}
\caption{\codename has higher accuracy than the state-of-the-art tools in detecting both storage and function collisions.}
\label{tab:accuracy-uschunt-dataset}
\end{table}

\para{Methodology}
We also compare the accuracy of \codename with USCHunt and CRUSH in detecting storage and function collisions.
For a fair comparison, we execute \codename, USCHunt, and CRUSH on the Smart Contract Sanctuary dataset~\cite{smart_contract_sanctuary}.
This dataset includes only contracts with available source code, enabling us to manually examine them and establish the ground truth data.

Specifically, within this dataset, USCHunt, CRUSH, and \codename identify 116, 102, and 55 storage collisions, respectively. 
Altogether, the three tools detect 206 unique storage collisions, of which we manually inspect the code.
In the case of function collisions, USCHunt identifies 300, while \codename reports 557, resulting in a total of 561 unique instances for manual verification.
Here, it is worth reminding that CRUSH does not detect function collisions.

\para{Results}
We report the collision detection accuracy in Table~\ref{tab:accuracy-uschunt-dataset}.
{\em Regarding storage collisions, \codename achieves an accuracy of 78.2\%}, surpassing both USCHunt and CRUSH, which each achieve an accuracy of 54.4\%.
USCHunt and CRUSH generate more false positives than \codename, albeit for different reasons.
Specifically, USCHunt mistakenly identifies variables with different names in separate contracts as collisions, often overlooking that one variable may serve as storage padding and is not exploitable.
\codename and CRUSH report a similar number of true positives (27 versus 26) thanks to their shared approach to detecting storage collisions. However, the higher accuracy of \codename over CRUSH is attributed to its effective identification of proxy smart contracts, wherein \codename precisely excludes library contracts.

{\em Regarding function collisions, \codename achieves an accuracy of 99.5\%, with no false positives and only three false negatives.}
In contrast, USCHunt has a notable lower accuracy of 53.3\%, with numerous false negatives due to the underlying Slither fails to identify proxy contracts.
Here, \codename also misses three function collisions due to runtime errors when emulating the EVM execution.

\section{Proxy Smart Contracts' Landscape}
\label{sec:landscape}

\begin{figure}[t!]
\begin{center}
  \includegraphics[width=0.45\textwidth]{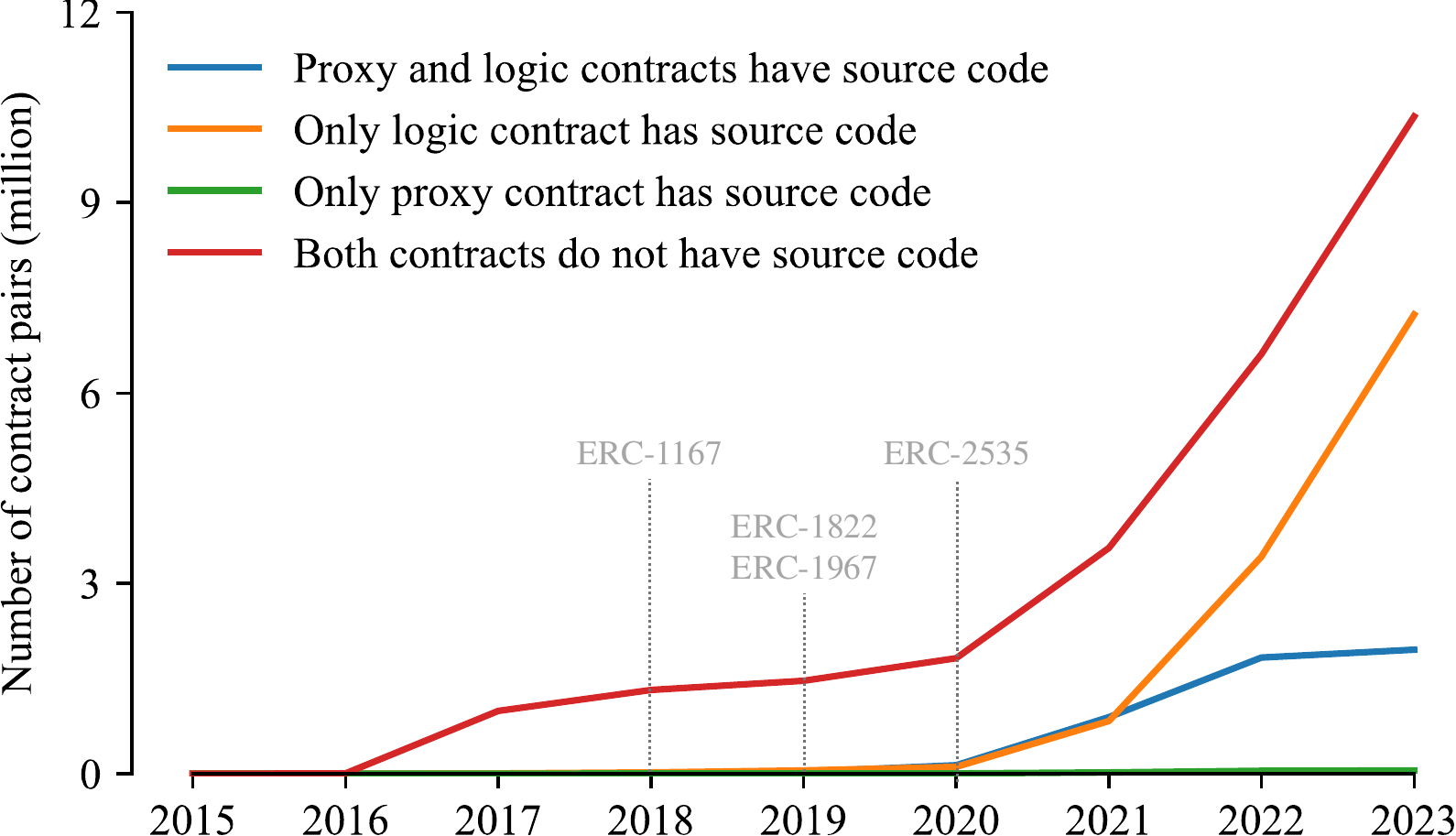}
  \caption{Accumulated number of pairs of proxy and logic contracts identified by \codename from 2015 to 2023. 
  In the vast majority of them, the proxy contracts only have bytecode available.
  }
  \label{fig:proxy-smart-contract-ratios}
\end{center}
\end{figure}

In this section, we present the comprehensive landscape of proxy smart contracts in Ethereum, resulting from \codename analyzing all alive contracts and detecting their function and storage collisions.

\subsection{Datasets}
\label{subsec:dataset}

We utilize \codename to examine all active smart contracts on the Ethereum network, deployed from the genesis block up to block 18473542 (the final block of October 2023).
We begin by querying the addresses and deployment blocks of all contracts from Google BigQuery~\cite{bigquery}.
Then, we retrieve their source code from EtherScan if available~\cite{etherscan}, and bytecode and storage states from a locally established Ethereum archive node~\cite{ethereum-archive-node}.
For efficient analysis, we assign the source code of a contract to all other contracts with the same bytecode hash. 
As previously shown in Figure~\ref{fig:smart-contract-ratios}, among roughly 64 million deployed smart contracts, 36 million are active as of October 2023.
\codename successfully analyzes 95.1\% of those active contracts without any errors from the EVM execution.

\subsection{Findings}
We present several findings regarding the growth of proxy contracts over the years, the number of found collisions, and the trends in their design patterns and deployments.

{\em First, over half of the active contracts are proxy or logic contracts, and the majority of proxy contracts do not publish their source code.}
We show the number of active proxy contracts identified by \codename in Figure~\ref{fig:proxy-smart-contract-ratios}.
As of October 2023, there are 19,599,317 proxy contracts, which represent 54.2\% of all contracts.
Of these, approximately 2 million are pairs of proxy and logic contracts with available source codes for both, as indicated by the blue line. Conversely, about 90\% of proxy contracts lack available source codes, as shown by the orange and red lines.
It should also be noted that in total, \codename reveals approximately 1.5 million {\em hidden} proxy contracts for which neither source code nor past transactions are available.

Furthermore, there has been a noticeable divergence in the growth trends of proxy contracts before and after 2020.
From 2015 to 2020, a total of only 2 million proxy contracts were deployed. 
However, the deployment of proxy contracts surged after 2020; for instance, 7.6 million proxy contracts were deployed in just the first ten months of 2023.
These figures closely track the historical adoption of the proxy pattern in the development of Ethereum smart contracts.
Particularly, before the first proxy-related EIPs~\cite{ethereum2023eip897, ethereum2023eip1167} in 2018, there was a clear demand for upgradeability and functionality cloning in smart contracts, demonstrated in about 1.3 million contracts already exploiting delegate calls. 
The period between 2018 and 2020 appears to be the testing phase, in which several EIPs~\cite{ethereum2023eip1167,ethereum2023eip1822,ethereum2023eip1967,ethereum2023eip2535} were proposed to standardize the proxy pattern, thus seeing a stable development in the number of proxy contracts.
Proxy contracts have become mainstream since 2020, especially in 2022 and 2023, when more than 93\% of contracts deployed are proxy contracts.


\begin{table}[t!]
\centering
\normalsize
\begin{tabular}{|c|c|c|}
    \hline
    \textbf{Year} & \textbf{Function collisions} & \textbf{Storage collisions}\\
    \hline
    \hline
    2017 & 24 & 0 \\
    2018 & 5,341 & 7 \\
    2019 & 16,136 & 37 \\
    2020 & 28,448 & 34 \\
    2021 & 705,801 & 725\\
    2022 & 808,493 & 2,082 \\
    2023 & 2,541 & 137\\
    \hline
    \hline
    \textit{Total} & \textit{1,566,784} &  \textit{3,022} \\
    \hline
\end{tabular}
\caption{Number of function and storage collisions detected by \codename. Notably, {\em 1,545,722 (or 98.7\%)} proxy contracts with function collisions are actually identical.}
\label{tab:number-of-collision}
\end{table}


{\em Second, \codename detects about 1.5 million function collisions, 98.7\% of which are duplicated contracts with the same code, and about 3 thousand exploitable storage collisions.}
We report the number of function and storage collisions found by \codename in Table~\ref{tab:number-of-collision}.
Specifically, starting from the 19.5 million pairs of proxy and logic contracts, \codename detects a total of $1,566,784$ pairs having function collisions and $3,022$ pairs having storage collisions. 
Notably, 98.7\% of the detected function collisions come from many proxy contracts duplicated from the {\em OwnableDelegateProxy}\footnote{\scriptsize\url{https://etherscan.io/address/0x0a08e6058eaaa847a1adb55b0a69b8469ea5a5b3}} contract.
In those cases, function collisions are caused by the {\tt proxyType()}, {\tt implementation()}, {\tt upgradeabilityOwner()} functions appearing in both proxy and logic contracts, possibly due to contract inheritance~\cite{zheng2021upgradable}. 

Regarding storage collisions, we identified 91 instances out of 3,022 where both proxy and logic contracts have their source code available.
This availability enables us to study their owners and the potential impacts of exploiting them. 
Specifically, we pinpointed 11 entities with smart contracts vulnerable to storage collisions, including Ape Finance, Compound, Convex, Curve, GolduckDAO, LeverFi, Poly, Polyhedra Network, Polymath, Tokeny, and Zora.
As of this writing, these entities manage stakes totaling 19 billion USD. 
However, it is important to acknowledge that they may manage stakes in other contracts that are not prone to such vulnerabilities.


\begin{figure}[t!]
     \centering
     \begin{subfigure}[b]{0.23\textwidth}
         \centering
         \includegraphics[width=\textwidth]{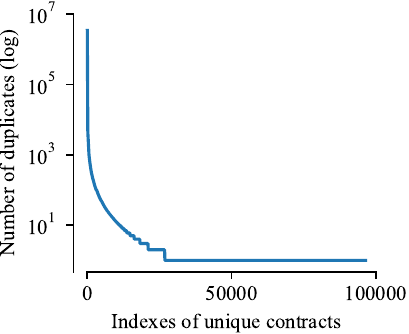}
         \caption{Proxy contracts}
         \label{subfig:unique-proxy}
     \end{subfigure}
     \begin{subfigure}[b]{0.23\textwidth}
         \centering
         \includegraphics[width=\textwidth]{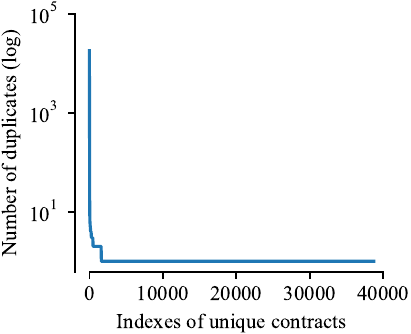}
         \caption{Logic contracts}
         \label{subfig:unique-logic}
     \end{subfigure}
        \caption{Most contracts are duplicates: only $96,420$ and $38,707$ unique proxy and logic contracts, respectively. 
        (\ref{subfig:unique-proxy}): three proxy contracts are duplicated more than 1 million times.
        (\ref{subfig:unique-logic}): two logic contracts have the same bytecode with more than ten thousand other contracts.
        }
        \label{fig:contract-uniqueness}
\end{figure}

{\em Third, we find the distributions of bytecode uniqueness are heavily skewed, with 42\% of proxy contracts duplicating from just three popular contracts.}
We highlight the number of unique proxy and logic contracts in Figure~\ref{fig:contract-uniqueness}.
Interestingly, while \codename identifies approximately 19.6 million proxy contracts and 70 thousand associated logic contracts, most of them are actually duplicates (i.e., having the same compiled bytecode) deployed at different addresses.
Particularly, Figure~\ref{fig:contract-uniqueness} reports only $96,420$ and $38,707$ unique proxy and logic contracts, respectively.
We see that the distributions of bytecode uniqueness are heavily skewed, in which a small number of contracts are duplicated significantly more than others. 
To be more specific, the three most popular proxy contracts have more than a million of duplicated contracts, and they are {\em CoinTool\_App}\footnote{\scriptsize\url{https://etherscan.io/address/0x95a3946104132973b00ec0a2f00f7cc2b67e751f}}, {\em XENTorrent}\footnote{\scriptsize\url{https://etherscan.io/address/0x4e488a5367daf86cfc71ea3b52ff72ca937efcf8}},  and {\em OwnableDelegateProxy} contracts.
We also notice that while most logic contracts are not duplicated frequently, there are two standout logic contracts\footnote{\scriptsize\url{https://etherscan.io/address/0xf17b1a1f68e1ddaa2e3285437b96ea28af2a2dc0}}\textsuperscript{,}\footnote{\scriptsize\url{https://etherscan.io/address/0xa471cd47769c3a788ad9c7b3d8350f195bf672bd}} having more than $10,000$ duplicates.
We conjecture that these contracts have source code, which renders cloning them uncomplicated and relates to the popularity of the non-fungible token marketplaces in recent years (e.g., {\em OwnableDelegateProxy} is a core component of the popular Wyvern protocol~\cite{wyvern}).
It is worth noting that all the duplicates of the popular proxy contracts above associate with the same logic contracts. 
For example, the {\em CoinTool\_App}  logic contract\footnote{\scriptsize\url{https://etherscan.io/address/0x0de8bf93da2f7eecb3d9169422413a9bef4ef628}} is referenced by almost 3.5 million proxy contracts that are duplicates of the {\em CoinTool\_App} proxy contract.
These findings indicate a widespread contract cloning practice that preserves the cloned contract's functionalities.
It may be, however, not ideal from the decentralization perspective, as potential bugs or vulnerabilities of the cloned contracts are also propagated, as noted in the previous paragraph or in existing work~\cite{yan2023bad}. 


\begin{table}[t!]
\centering
\normalsize
\begin{tabular}{|l|c|c|}
    \cline{2-3}
    \multicolumn{1}{c|}{} & \textbf{\# Contracts} & \textbf{Ratio}\\
    \hline
    \hline
    EIP-1167~\cite{ethereum2023eip1167} & $17,453,264$ & $89.05\%$ \\
    \hline
    EIP-1822~\cite{ethereum2023eip1822} & $22,789$ & $0.12\%$ \\
    \hline
    EIP-1967~\cite{ethereum2023eip1967} & $196,688$ & $1.00\%$ \\
    \hline
    Others & $1,926,576$ & $9.83\%$\\
  \hline
\end{tabular}
\caption{The number of proxy contracts following certain design standards. The vast majority of proxy contracts follow the minimal design (EIP-1167). \codename misses only a few hundred of the diamond proxy contracts (EIP-2535).}
\label{tab:proxy-smart-contract-compliance-ratios}
\end{table}

{\em Fourth, the minimal proxy design dominates the standard proxy contracts while a non-negligible portion of non-standard proxy contracts exists.}
We present the distribution of proxy contracts' design patterns in Table~\ref{tab:proxy-smart-contract-compliance-ratios}.
Notably, most of them (89.05\%) adhere to the minimal design standard~\cite{ethereum2023eip1167}, which includes only the delegate call in the fallback function and a hard-coded logic contract address in the bytecode.
These minimal proxy contracts are not vulnerable to function and storage collisions due to the absence of variable and function declarations.
We further categorize proxy contracts into other standards based on the locations storing their logic contracts' addresses.
In particular, $22,789$ contracts are categorized under the EIP-1822 standard (Universal Upgradeable Proxy Standard) as they utilize a specific storage slot derived from the {\tt Keccak-256("PROXIABLE")} hash.
Similarly, $196,688$ contracts follow the EIP-1967 standard because they store logic contract addresses in a slot derived from the {\tt Keccak-256} hash of {\tt ("eip1967.proxy.implementation")}.
We also note that there are $9.83\%$ of the proxy contracts storing their logic contract's addresses in the storage without conforming to any known design patterns.

\begin{figure}[t!]
\begin{center}
  \includegraphics[width=0.48\textwidth]{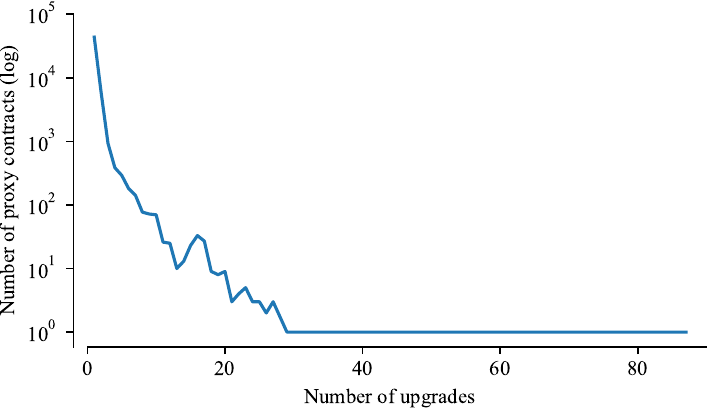}
  \caption{Number of upgrades for logic contracts in log scale. Most proxy contracts (99.7\%) have not upgraded to a newer version of logic contracts.}
  \label{fig:number-of-upgrades}
\end{center}
\end{figure}

{\em Last, we find that most proxy contracts have not upgraded to a newer version of logic contracts.}
We show the number of contract upgrade events in which the logic contract's address is updated in Figure~\ref{fig:number-of-upgrades}.
We find that the number of upgraded smart contracts is insignificant, e.g., only $51,925$ proxy contracts have upgraded their logic implementations throughout history.
The majority of these contracts also upgrade only a few times (i.e., having only 1.32 associated logic contracts on average).
We also observe that upgrade events are infrequent; only $68,804$ upgrading events ever occurred, meaning, on average, an upgrade happens only once per 200 Ethereum blocks.
From the security point of view, such an infrequent upgrade may render proxy contracts less prone to storage collisions, which often arise during contract upgrades. 
 
\section{Discussion} 
\label{sec:discussion}
In this section, we discuss several key points. 
We first describe the limitations of \codename and explain why they do not undermine \codename's contributions (\S\ref{subsec:limitations}).
Based on the discussed limitations, we outline a few potential follow-up works for \codename (\S\ref{subsec:future-work}). 
We finally acknowledge the ethical considerations in this paper (\S\ref{subsec:ethical}).


\subsection{Limitations}
\label{subsec:limitations}

When analyzing proxy contracts, \codename misses the contracts following the Diamonds, Multi-Facet Proxy design pattern~\cite{ethereum2023eip2535}, in which only the function signatures registered by the contract owner can trigger the delegate calls in the fallback function.
Unfortunately, \codename can currently only send randomly generated call data during EVM emulation and, thus, cannot detect these diamond contracts.


Another limitation of the system is the occurrence of runtime errors, which are relatively low (e.g., at 4.9\%) during the emulation of EVM execution of smart contracts (cf. Section~\ref{subsec:dataset}).
Additionally, EVM emulation may inevitably yield results that differ from actual contract execution, although the extent of these discrepancies is not known.



\subsection{Future Work}
\label{subsec:future-work}

Future work for \codename includes identifying proxy contracts that follow the diamond design pattern~\cite{ethereum2023eip2535}. 
A potential solution involves extracting all registered functions from past transactions (similar to CRUSH~\cite{ruaro2024not}) and utilizing them to generate call data.
To detect contracts with source code available, \codename may employ a static analysis approach like USCHunt~\cite{bodell2023proxy}, combining with the information of the slot storing logic contract's address (i.e., {\tt Keccak-256("diamond.standard.diamond.storage")}).



\codename can also be extended to analyze proxy smart contracts beyond Ethereum blockchain.
Similar to USCHunt~\cite{bodell2023proxy}, \codename may apply to several other blockchains, such as Arbitrum, Avalanche, Binance Smart Chain, Celo, Fantom, Optimism, and Polygon.





\subsection{Ethical Considerations}
\label{subsec:ethical}

This paper does not raise any ethical issues. 
Our evaluation uses the already available datasets. 
Our interaction with the Ethereum production network is only to retrieve real-world data and does not affect any other parties.

\section{Related Work}
\label{sec:related-work}

We consider related work that studies the same target of proxy smart contracts (\S\ref{subsec:related-proxy}), discusses contract upgradability (\S\ref{subsec:related-upgradeability}), or performs analysis on smart contracts (\S\ref{subsec:related-analyzer}).

\subsection{Finding Collisions in Proxy Smart Contracts}
\label{subsec:related-proxy}

Several tools aim to detect proxy smart contracts and their collision issues; see Table~\ref{tab:coverage} for an overview.

Slither, a static analysis framework, examines the source code of contracts to determine if they are proxy contracts~\cite{feist2019slither}. 
However, Slither's proxy detection relies on keyword searches, such as "proxy" or "delegatecall," which may lead to a high rate of false positives. 
Additionally, in contrast to \codename, Slither does not automatically identify associated logic contracts for a given proxy contract.

Etherscan is a widely recognized web-based explorer for the Ethereum blockchain, featuring an integrated proxy contract verification tool~\cite{etherscan}. 
This tool identifies contracts with the \delegate opcode as proxy contracts, a result that Etherscan admits may lead to numerous false positives~\cite{etherscan-proxy}. 
\codename applies a similar initial filtering process for proxy contracts and then conducts dynamic analysis, resulting in more precise detection.

USCHunt~\cite{bodell2023proxy} employs static analysis based on Slither to detect proxy contracts with published source code, and their security vulnerabilities, such as collisions, on eight blockchains, including Ethereum.
\codename focuses on improving the detection of proxy contracts and collision issues, specifically on the Ethereum blockchain, which has shown to be more efficient, inclusive, and precise than USCHunt throughout this paper.

Salehi et al.~\cite{salehi2022not} studied the ownership of upgradability in smart contracts, i.e., finding out who can upgrade the proxy contracts.
Similar to \codename, this work also performs dynamic analysis on smart contracts' bytecode, thus covering more contracts than USCHunt. 
Unlike \codename, however, the analysis here involves replaying past transactions to the contracts under tests, thus limiting the effective analysis to only contracts with many transactions. 

CRUSH is a newly developed automation tool that detects storage collisions and generates verified exploits~\cite{ruaro2024not}.
The CRUSH engine is also employed by \codename to identify storage collisions, particularly for proxy contracts without source code. Like Salehi et al., CRUSH depends on historical transactions to locate proxy contracts, thus missing out on millions of hidden contracts. 
Additionally, unlike \codename, CRUSH is not equipped to detect function collisions.

\subsection{Upgradeability in Smart Contracts}
\label{subsec:related-upgradeability}

The upgradeability in smart contracts has been discussed extensively in several EIPs~\cite{ethereum2023eip1167, ethereum2023eip1822, ethereum2023eip1967, ethereum2023eip2535}, technical blog posts~\cite{openzeppelin, trail_of_bits}, and numerous studies~\cite{frowis2022not,zheng2021upgradable,klinger2020upgradeability,rodler2021evmpatch}.
These works focus on designing new proxy patterns to enable upgrading smart contracts at scale or studying the status quo of upgradable contracts. 
Our work also studies upgradable contracts in Ethereum, which is a subset of proxy smart contracts. 
We further reveal that upgrading events are actually rare, and functionality cloning is a more popular usage of proxy contracts. 
Our work, thus, provides an additional discussion to the existing literature on upgradeability in smart contracts.

\subsection{Smart Contracts Analyzers}
\label{subsec:related-analyzer}

The security of smart contracts is a well-studied research topic, as smart contracts are known to have multiple vulnerabilities~\cite{atzei2017survey}.
Many smart contract analyzers have been proposed in recent years to detect these vulnerabilities.
Commonly, they can be categorized into static analyzers and dynamic analyzers based on their overall approach. 
Particularly, static analyzers detect smart contract vulnerabilities by inspecting their source code or bytecode, using techniques such as information flow analysis (e.g., Slither~\cite{feist2019slither}, MadMax~\cite{grech2018madmax}) or symbolic execution (e.g., Oyente~\cite{luu2016making}, MantiCore~\cite{mossberg2019manticore}, Securify~\cite{tsankov2018securify}, teEther~\cite{krupp2018teether}, Mythril~\cite{mythril}, Zeus~\cite{kalra2018zeus}, Osiris~\cite{torres2018osiris}).
On the other hand, dynamic analyzers execute the test smart contracts and observe the behaviors of vulnerable ones, using fuzzing (e.g., ReGuard~\cite{liu2018reguard}, ContractFuzzer~\cite{jiang2018contractfuzzer}, Confuzzius~\cite{torres2021confuzzius}, sFuzz~\cite{nguyen2020sfuzz}, Harvey~\cite{wustholz2020harvey}) or validation (e.g., MAIAN~\cite{nikolic2018finding}, Sereum~\cite{rodler2018sereum}, SODA~\cite{chen2020soda}, ESCORT~\cite{sendner2023smarter}).
We refer to a recent survey by Kushwaha et al.~\cite{kushwaha2022ethereum} for a more comprehensive review of such existing analysis tools.
This paper proposes \codename, a new hybrid contract analyzer focusing specifically on the collision issues of Ethereum smart contracts.
\section{Conclusion}
\label{sec:conclusion}

While the proxy design pattern facilitates upgrading smart contracts in Ethereum, it also introduces potential security risks due to function and storage collisions. 
Previous efforts to detect these collisions have been inadequate, mainly due to the limited coverage of contracts under test, especially for the hidden ones without source code or transaction history.
To bridge this gap, we have created \codename, a novel automated tool that efficiently, effectively, and accurately identifies hidden proxy contracts and their associated collision vulnerabilities.
Utilizing \codename, we analyzed all Ethereum smart contracts and discovered that half are associated with the proxy pattern, and millions of them are vulnerable to collisions.
Observing proxy smart contracts' evolution suggests an increased adoption of the proxy pattern in the coming years. 
We thus believe our work significantly contributes to securing proxy contracts in this anticipated future.

\bibliographystyle{ACM-Reference-Format}
\bibliography{main}


\end{document}